\documentclass[10pt,amsmath,amssymb,latexsym]{article}
\usepackage{graphicx}  
\usepackage{dcolumn}   
\usepackage{bm}        
\usepackage{verbatim}  
\usepackage{amsmath}
\usepackage{leftidx}
\usepackage{tikz}
\usepackage{graphicx}

\newtheorem{myProperty}{Property}

\newcommand{\state}[1]{\left|{#1}\right>}
\newcommand{\stateBasis}[2]{\left|{#1}\right>_{#2}}

\newcommand{\braBasis}[2]{\left<{#1}\right|_{#2}}

\newcommand{\braketBasis}[3]{\leftidx{_{#3}}{\left<{#1}|{#2}\right>}_{#3}}

\newcommand{\PartTr}[2]{\textrm{Tr}_{#2}\left[{#1}\right]}

\begin{document}

\title{
Quantum mechanics allows undetectable inconsistencies in witnessed events
}
\author{Hitoshi Inamori\\
\small\it Soci\'et\'e G\'en\'erale\\
\small\it Boulevard Franck Kupka, 92800 Puteaux, France\\
}

\bigskip

\date{\today}

\maketitle

\begin{abstract}
Quantum mechanics, devoid of any additional assumption, does not give any theoretical constraint on the projection basis to be used for the measurement process. It is shown in this paper that it does neither allow any physical means for an experimenter to determine which measurement bases have been used by another experimenter. As a consequence, quantum mechanics allows a situation in which two experimenters witness incoherent stories without being able to detect such incoherence, even if they are allowed to communicate freely by exchanging iterative and bilateral messages.
 
\bigskip


\end{abstract}

\section{Introduction}
Bob is doing an experiment. He is performing a measurement on a quantum system, which is initially in a superposition of states A and B. The outcomes A or B have equal probability of occurring. Bob runs the experiment and Alice gets a call from Bob telling he got result A. 
What can Alice say about what Bob has actually witnessed during the experiment?

The question here is not about Bob's honesty, but whether it is possible that Bob witnessed a result unrelated to what Alice heard from him. At first this seems absurd. Of course, Bob could not have communicated the result A if he had not witnessed that result!   

However, the real question here is about the coherence between what Alice heard from Bob on Alice's side, and what Bob actually observed on his side. 

Considering the laws of quantum mechanics exempt from any further assumptions, we see that the above question is not a trivial one, once we realize that Bob's subjective experience cannot be proven to be the cause, nor the result, of Bob's behaviour observed from the outside.

This note presents why the actual observations made by Bob may turn out to be at odds with what Alice heard from Bob. Moreover, we present why there is no physical mean for Alice to detect the possible incoherences between the observations made by Alice and Bob in general. 
This may lead to a situation in which Alice's perspective of the world and Bob's perspective of the world may be incompatible, even if Alice sees Bob agreeing with her perspective of the world, and Bob sees Alice agreeing with his perspective of the world.


\section{A very simple example}  
Here is the simplest and naive illustration of what can lead to the situation presented in the Introduction.
In Figure~\ref{Fig1}, Bob performs a measurement on a qubit, called the ``source'', initially in a state $\state{\psi}$,  and communicates the measurement outcome to Alice.

\setlength{\unitlength}{0.025 in}
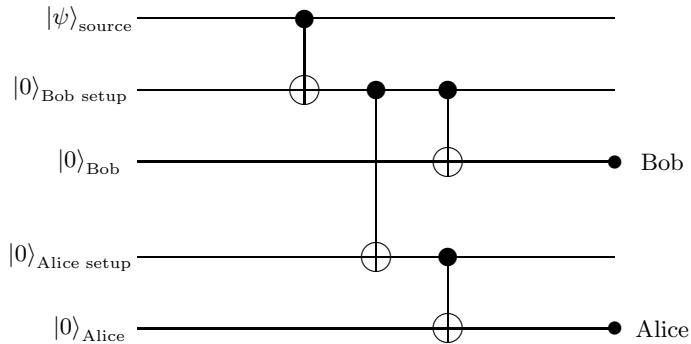
\begin{figure}[!h]
\centering

\begin{picture}(130,80)
\put(5,75){\makebox(0,0){\small $\stateBasis{\psi}{\text{source}}$}}
\put(15,75){\line(1,0){100}}
\put(50,75){\circle*{4}} 
\put(50,75){\line(0,-1){18}}

\put(1,60){\makebox(0,0){\small $\stateBasis{0}{\text{Bob setup}}$}}
\put(15,60){\line(1,0){100}}
\put(50,60){\circle{6}}
\put(65,60){\circle*{4}}
\put(65,60){\line(0,-1){38}}
\put(80,60){\circle*{4}}
\put(80,60){\line(0,-1){18}}

\put(5,45){\makebox(0,0){\small $\stateBasis{0}{\text{Bob}}$}}
\put(15,45){\line(1,0){100}}
\put(80,45){\circle{6}}
\put(115,45){\circle*{3}}
\put(125,45){\makebox(0,0){\small Bob}}

\put(1,25){\makebox(0,0){\small $\stateBasis{0}{\text{Alice setup}}$}}
\put(15,25){\line(1,0){100}}
\put(65,25){\circle{6}}
\put(80,25){\circle*{4}}
\put(80,25){\line(0,-1){18}}

\put(5,10){\makebox(0,0){\small $\stateBasis{0}{\text{Alice}}$}}
\put(15,10){\line(1,0){100}}
\put(80,10){\circle{6}}
\put(115,10){\circle*{3}}
\put(125,10){\makebox(0,0){\small Alice}}

\end{picture}
\caption{Alice receives a signal from Bob telling Bob's measurement outcome}\label{Fig1}
\end{figure}

The source qubit is entangled with Bob's setup that propagates this entanglement to Alice's setup (second Controlled-NOT from the left). Bob's setup here represents all the mechanism that allows Bob to perform a measurement on the source quantum system and to communicate the outcome of that measurement to Alice. Alice's setup represents all the mechanism that allows Alice to receive the communication from Bob.

Bob observes the state of his setup by conducting a Positive-Operator Valued Measure measurement (POVM) on it (represented by an entanglement with the probe representing Bob, initially in state $\stateBasis{0}{\text{Bob}}$, and conducting a Von Newman projective measurement on this probe). Alice does the same on her setup. 

The reader may object that the communication from Bob to Alice is necessarily a result of the POVM performed by Bob on his qubit. It is important to realize that, although such view is intuitive, it is not necessarily true nor proven. Certain experiments in neuroscience~\cite{Sirigu09} demonstrate, on the contrary, that having the subjective feeling of executing a gesture has no causal relationship with the execution of the gesture itself. Such experimental results encourage us to adopt a more general description in which Bob's subjective experience is not causally related with Bob's action as observed from the outside.

The setup described in Figure~\ref{Fig1} would imply a perfect correlation between the measurement outcome of Bob's qubit and the measurement outcome of Alice's qubit: Bob and Alice would experience the same outcome.

\begin{figure}[!h]
\centering

\begin{picture}(130,80)
\put(5,75){\makebox(0,0){\small $\stateBasis{\psi}{\text{source}}$}}
\put(15,75){\line(1,0){100}}
\put(50,75){\circle*{4}} 
\put(50,75){\line(0,-1){18}}

\put(1,60){\makebox(0,0){\small $\stateBasis{0}{\text{Bob setup}}$}}
\put(15,60){\line(1,0){100}}
\put(50,60){\circle{6}}
\put(65,60){\circle*{4}}
\put(65,60){\line(0,-1){38}}
\put(80,60){\circle*{4}}
\put(80,60){\line(0,-1){18}}

\put(5,45){\makebox(0,0){\small $\stateBasis{0}{\text{Bob}}$}}
\put(15,45){\line(1,0){80}}
\put(80,45){\circle{6}}
\put(115,45){\circle*{3}}
\put(95,40){\framebox(10,10){$H$}}
\put(105,45){\line(1,0){10}}

\put(125,45){\makebox(0,0){\small Bob}}

\put(1,25){\makebox(0,0){\small $\stateBasis{0}{\text{Alice setup}}$}}
\put(15,25){\line(1,0){100}}
\put(65,25){\circle{6}}
\put(80,25){\circle*{4}}
\put(80,25){\line(0,-1){18}}

\put(5,10){\makebox(0,0){\small $\stateBasis{0}{\text{Alice}}$}}
\put(15,10){\line(1,0){100}}
\put(80,10){\circle{6}}
\put(115,10){\circle*{3}}
\put(125,10){\makebox(0,0){\small Alice}}

\end{picture}
\caption{Alice receives a signal from Bob telling Bob's measurement outcome. However Bob's experience is based on a measurement performed in an alternate basis.}\label{Fig2}
\end{figure}
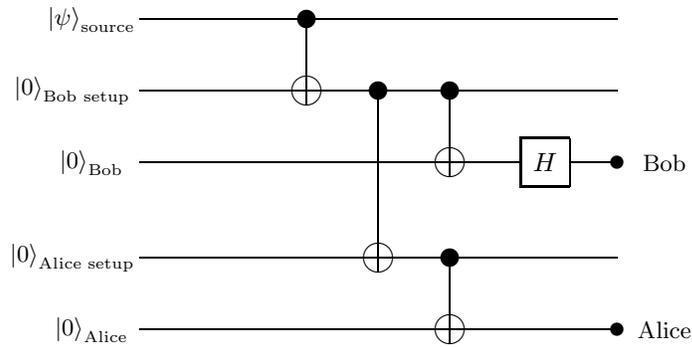

However, there is no way for Alice to guarantee that Bob performs the measurement as described in Figure~\ref{Fig1}. In particular, Bob's qubit could be subjected to another unitary transformation before the measurement, such as the Hadamard transform added in Figure~\ref{Fig2}.

This change of basis is not observable from Alice, as the density matrix of the system available to Alice is not disrupted by any transformation made on Bob's side, as long as this transformation is made after the last interaction with the system available to Alice. Note that other types of modifications on Bob's quantum channel are also possible without Alice noticing, but in this example we will focus on the selection of the measurement basis for Bob's measurement.

This is a very simple illustration of the following phenomenon:  Alice gets a signal from Bob's setup telling that it got a certain measurement outcome. Yet at Bob's level, the observation he experiences may be completely uncorrelated with what Alice heard from Bob's setup.

\section{Second example: incoherence between witnessed correlations}

The above possible inconsistency between Alice's and Bob's subjective experience can be easily extended to the witnessed relationships between events observed by Alice and Bob. It could lead to a situation in which Alice gets a message from Bob's setup stating the presence of a correlation between two observables, while Bob does actually not experience such correlation on his side (or vice versa).

\begin{figure}[!h]
\centering

\begin{picture}(130,120)

\put(3,110){\makebox(0,0){\small $\stateBasis{0}{\text{source}_1}$}}
\put(15,110){\line(1,0){10}}
\put(25,105){\framebox(10,10){$H$}}
\put(35,110){\line(1,0){80}}
\put(45,110){\circle*{4}}
\put(45,110){\line(0,-1){13}}
\put(55,110){\circle*{4}}
\put(55,110){\line(0,-1){28}}
\put(65,100){\circle*{4}}
\put(65,100){\line(0,-1){28}}

\put(3,100){\makebox(0,0){\small $\stateBasis{0}{\text{source}_2}$}}
\put(15,100){\line(1,0){100}}
\put(45,100){\circle{6}}

\put(1,85){\makebox(0,0){\small $\stateBasis{0}{\text{Bob setup}_1}$}}
\put(15,85){\line(1,0){100}}
\put(55,85){\circle{6}}
\put(95,85){\circle*{4}}
\put(95,85){\line(0,-1){23}}
\put(75,85){\circle*{4}}
\put(75,85){\line(0,-1){48}}
\put(85,75){\circle*{4}}
\put(85,75){\line(0,-1){48}}

\put(1,75){\makebox(0,0){\small $\stateBasis{0}{\text{Bob setup}_2}$}}
\put(15,75){\line(1,0){100}}
\put(65,75){\circle{6}}
\put(105,75){\circle*{4}}
\put(105,75){\line(0,-1){23}}

\put(3,65){\makebox(0,0){\small $\stateBasis{0}{\text{Bob}_1}$}}
\put(15,65){\line(1,0){100}}
\put(95,65){\circle{6}}
\put(115,65){\circle*{3}}
\put(125,65){\makebox(0,0){\small $\text{Bob}_1$}}

\put(3,55){\makebox(0,0){\small $\stateBasis{0}{\text{Bob}_2}$}}
\put(15,55){\line(1,0){100}}
\put(105,55){\circle{6}}
\put(115,55){\circle*{3}}
\put(125,55){\makebox(0,0){\small $\text{Bob}_2$}}

\put(1,40){\makebox(0,0){\small $\stateBasis{0}{\text{Alice setup}_1}$}}
\put(15,40){\line(1,0){100}}
\put(75,40){\circle{6}}
\put(95,40){\circle*{4}}
\put(95,40){\line(0,-1){23}}

\put(1,30){\makebox(0,0){\small $\stateBasis{0}{\text{Alice setup}_2}$}}
\put(15,30){\line(1,0){100}}
\put(85,30){\circle{6}}
\put(105,30){\circle*{4}}
\put(105,30){\line(0,-1){23}}

\put(3,20){\makebox(0,0){\small $\stateBasis{0}{\text{Alice}_1}$}}
\put(15,20){\line(1,0){100}}
\put(95,20){\circle{6}}
\put(115,20){\circle*{3}}
\put(125,20){\makebox(0,0){\small $\text{Alice}_1$}}

\put(3,10){\makebox(0,0){\small $\stateBasis{0}{\text{Alice}_1}$}}
\put(15,10){\line(1,0){100}}
\put(105,10){\circle{6}}
\put(115,10){\circle*{3}}
\put(125,10){\makebox(0,0){\small $\text{Alice}_2$}}

\end{picture}
\caption{Bob observes two entangled qubits and communicates the two-bit results to Alice.}\label{Fig3}
\end{figure}
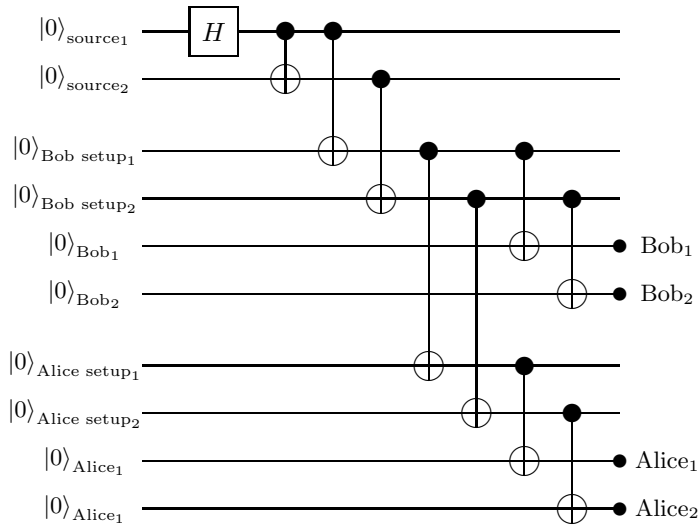

Consider an experience similar to the one in the previous section, but in this case Bob's setup performs a measurement on a qubit pair and reports the measurement outcome to Alice. Bob performs a POVM on his setup and witnesses the two-bit outcome. So does Alice on her side (Figure~\ref{Fig3}).

This setup should again imply a perfect correlation between the measurement outcomes observed by Alice and Bob. In particular, Alice and Bob should witness the same perfect correlation between the two bits they observe: the first bit of the pair is zero and the second bit of the pair is zero, or the first bit is one and the second bit is one.

Suppose now that the measurement basis for Bob's two-qubit system $(\text{Bob}_1, \text{Bob}_2)$ is altered, by for instance, applying the Hadamard gate on $\text{Bob}_2$ (and only on $\text{Bob}_2$)  just before Bob's measurement. Such basis change will eliminate the correlation between  the pair of bits on Bob's side, while Alice will continue seeing perfect correlation.

As correlation is a necessary -- albeit not sufficient -- condition for causality (causality being a much more subjective notion than correlation), this implies that Alice and Bob may not see the same causal relationships between observed events.

Put in a wider context, such effect could lead to a difference in the relationships between events observed by Alice and Bob, i.e. the implied physical laws that Alice and Bob are witnessing. 
Depending on the measurement bases chosen by Alice and Bob, Alice could witness correlation between some dependent physical values while Bob could see no such correlation. More importantly, there would be no means for Alice to detect that Bob is not experiencing the same correlations as seen by her, as the messages she gets from Bob are in line with her perspective.  Indeed, the change of the measurement basis for Bob's qubits have no impact on the density matrix available to Alice. This means that there exists no physical means for Alice to guess the basis chosen for Bob's measurement on his own pair of qubits.

\section{General case}
\newsavebox{\bigschema}
\savebox{\bigschema}
  (200,100)[bl]{
\put(3,0){\makebox(0,0){\small $t_0$}}
\put(185,0){\makebox(0,0){\small $t$}}

\put(3,50){\makebox(0,0){\small $\rho_0$}}
\put(10,05){\line(0,1){90}}
\put(10,05){\line(1,0){5}}
\put(10,95){\line(1,0){5}}

\put(15,90){\line(1,0){15}}
\put(15,74){\line(1,0){15}}
\put(15,58){\line(1,0){15}}
\put(15,42){\line(1,0){15}}
\put(15,26){\line(1,0){15}}
\put(15,10){\line(1,0){15}}

\put(30,5){\framebox(20,90){$U_0$}}

\put(50,90){\line(1,0){25}}
\put(85,95){\makebox(0,0){\scriptsize POVM giving $\beta_1$}}
\put(75,90){\circle*{4}}
\put(75,89.5){\line(1,0){4.5}}
\put(75,90.5){\line(1,0){5.5}}
\put(79.5,89.5){\line(0,-1){10.5}}
\put(80.5,90.5){\line(0,-1){11.5}}

\put(70,5){\framebox(20,74){$U_{\beta_1}$}}

\put(110,21){\framebox(20,58){$U_{\alpha_1}$}}

\put(150,21){\framebox(20,42){$U_{\beta_2}$}}

\put(50,74){\line(1,0){20}}
\put(90,74){\line(1,0){20}}
\put(130,74){\line(1,0){25}}
\put(165,79){\makebox(0,0){\scriptsize POVM giving $\beta_2$}}
\put(155,74){\circle*{4}}
\put(155,73.5){\line(1,0){4.5}}
\put(155,74.5){\line(1,0){5.5}}
\put(159.5,73.5){\line(0,-1){10.5}}
\put(160.5,74.5){\line(0,-1){11.5}}

\put(50,58){\line(1,0){20}}
\put(90,58){\line(1,0){20}}
\put(130,58){\line(1,0){20}}
\put(170,58){\line(1,0){10}}
\put(187,58){\makebox(0,0){\small Bob}}

\put(50,42){\line(1,0){20}}
\put(90,42){\line(1,0){20}}
\put(130,42){\line(1,0){20}}
\put(170,42){\line(1,0){10}}

\put(50,26){\line(1,0){20}}
\put(90,26){\line(1,0){20}}
\put(130,26){\line(1,0){20}}
\put(170,26){\line(1,0){10}}
\put(187,26){\makebox(0,0){\small Alice}}

\put(50,10){\line(1,0){20}}
\put(90,10){\line(1,0){25}}
\put(125,05){\makebox(0,0){\scriptsize POVM giving $\alpha_1$}}
\put(115,10){\circle*{4}}
\put(115,10.5){\line(1,0){4.5}}
\put(115,9.5){\line(1,0){5.5}}
\put(119.5,10.5){\line(0,1){10.5}}
\put(120.5,9.5){\line(0,1){11.5}}
}

\newsavebox{\equivalent}
\savebox{\equivalent}
  (100,100)[bl]{
\put(3,0){\makebox(0,0){\small $t_0$}}
\put(82,0){\makebox(0,0){\small $t$}}

\put(3,50){\makebox(0,0){\small $\rho_0$}}
\put(10,05){\line(0,1){90}}
\put(10,05){\line(1,0){5}}
\put(10,95){\line(1,0){5}}

\put(15,75){\line(1,0){15}}
\put(15,50){\line(1,0){15}}
\put(15,25){\line(1,0){15}}

\put(30,5){\framebox(30,90){$U_{t,t_0}$}}

\put(60,75){\line(1,0){15}}
\put(60,50){\line(1,0){15}}
\put(60,25){\line(1,0){15}}

\put(82,75){\makebox(0,0){\small Bob}}

\put(82,25){\makebox(0,0){\small Alice}}
}

In the case studies above, we have analyzed oversimplified experiments in which the interaction between Alice and Bob was done through single qubits, and in which the communication between Alice and Bob was limited to a single, one-way communications from Bob to Alice. One could argue that a more realistic representation of the experiment would involve far more degrees of freedom. One could also add an interactive communication between Alice and Bob. This could give some credibility to Alice that Bob is actually observing what Alice hears from him through exchanged messages.
 
\begin{figure}[!h]
\centering

\begin{picture}(200,100)

\put(0,0){\scalebox{0.95}{\usebox{\bigschema}}}
 \end{picture}
\caption{Example of an interactive measurement process. }\label{Fig4}
\end{figure}
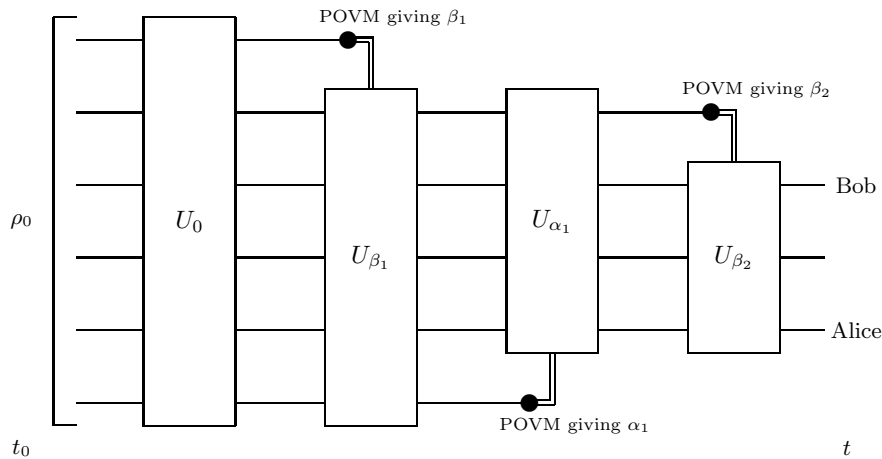

Such interactive process can be represented as in Figure~\ref{Fig4}. Single lines represent quantum channel (of arbitrary dimension) and double lines represent classical information flow.

The whole quantum system starts in an initial state described by a density operator $\rho_0$ at time $t_0$, and we are interested in the state of the experiment  at some time $t\geq t_0$, time at which Alice and Bob can perform a last measurement on the subsystem that is available to each of them.

Whatever the size and the complexity of the setup, according to quantum mechanics, the setup undergoes a unitary transformation during time intervals in which neither experimenter performs a measurement. The system undergoes a first unitary transformation denoted by $U_0$. 

In this example, we assume that the first measurement is performed by Bob. It can be any POVM that Bob performs on a subsystem that is available to him, yielding an outcome denoted $\beta_1$. Bob can act on the subsequent evolution of the rest of the system depending on this outcome. For instance, Bob could send a message with the measurement outcome to Alice, write down the outcome on a piece of paper (the piece of paper being part of the quantum system), or could switch a button of the experimental setup depending on the outcome. We can represent this completely by having the unitary transformation after this measurement, $U_{\beta_1}$, being explicitly dependent on the outcome $\beta_1$ of this first measurement. Note that if Alice gets any information about the outcome $\beta_1$, it is through the transformation $U_{\beta_1}$, i.e. there must be a physical process conveying this information to Alice. 
Note also that any remembrance by Bob of the outcome $\beta_1$ (such as Bob writing down the outcome $\beta_1$ on a piece of paper) is also done through the transformation  $U_{\beta_1}$ on the phsyical system (in this example, the piece of paper) that will be available to Bob subsequently.

We suppose in this example that this is followed by a measurement performed by Alice. Again this can be any POVM applied on a subsystem available to Alice, yielding an outcome denoted $\alpha_1$. Like Bob, Alice acts on the remaining system contditional on this outcome $\alpha_1$, and this is described by the unitary operator $U_{\alpha_1}$. This interactive process between Alice and Bob goes on, until the time $t$ of interest.

Intuitively, the interactive nature of this experimental protocol seems to ensure that what Alice is witnessing and what Bob is witnessing are matched: what Bob is actually witnessing should correspond to what Alice understands from Bob's communication, and vice versa. Indeed, Alice hears from Bob what he observes, and Alice can ask questions interactively to verify the likelihood that Bob is actually seeing what Alice hears from him.

We are going to show that this intuition is not necessarily correct, however.

Let's first remove the technical complexity arising from the intermediary POVM's. It is well known~\cite{Bergou, KurtJacobs}, that any POVM, performed on a target quantum system, can be described as: (1)~a unitary transformation performed on a joint system, in which the target system is entangled with a probe quantum system, followed by (2)~a von Neumann projective measurement on that probe system. 

By enlarging, if necessary, the quantum system to include these probe systems, we can therefore assume, without loss of generality, that all the intermediary measurements in the interactive process above are in fact von Neumann projective measurements. 

The interactive process is therefore a sequence of projective measurements followed by conditional unitary transformations, until time $t$. It is known that such a sequence of projective measurements followed by conditional unitary transformations can be represented as a single unitary transformation~\cite{KurtJacobs}. In our case, we can show that, seen from Alice and Bob, this interactive process up to $t$ is not distinguishable from a single unitary process from the initial time $t_0$ until time $t$, time at which each party can perform a measurement on his or her subsystem. To see this, take any sequence in which a projective measurement is done on a subsystem $M$, followed by a conditional unitary transformation $U_m$ on the remaining subsystem $S$. The subsystem $M$ contains in particular the subsystems available to Alice and Bob. The subscript $m$ stands for any possible outcome from the measurement on $M$. Let $\{\stateBasis{m}{M}\}_m$ be the orthonormal basis representing the projective measurement on $M$, and $H_M$ represent the Hilbert space corresponding to $M$.

The above sequence maps any initial state $\stateBasis{\psi}{M}\otimes\stateBasis{\phi}{S}\in H_M\otimes H_S$,  to a mixture of states in which we get the state $\stateBasis{m}{M} \otimes U_m\stateBasis{\phi}{S}$ with the probability $|\braketBasis{m}{\psi}{M}|^2$. The state of the subsystem $S$ is obtained by tracing over $H_M$ the density matrix describing this mixture, which gives

\begin{eqnarray}
\rho_{S} &=& \PartTr{\sum_m |\braketBasis{m}{\psi}{M}|^2\stateBasis{m}{M}\braBasis{m}{M}\otimes U_m\stateBasis{\phi}{S}\braBasis{\phi}{S} U_m^\dag}{H_M} \\
&=& \sum_m |\braketBasis{m}{\psi}{M}|^2\, U_m\stateBasis{\phi}{S}\braBasis{\phi}{S} U_m^\dag.
\end{eqnarray}
 
Using a technique similar to what is used in~\cite{GriffithsNiu,KurtJacobs}, introduce the unitary operator  $W=\sum_m \stateBasis{m}{M}\braBasis{m}{M}\otimes U_m$ acting on $H_M\otimes H_S$. If we apply the unitary operator $W$ on $\stateBasis{\psi}{M}\otimes\stateBasis{\phi}{S}$, and if we look at the state of the subsystem $S$ by tracing over $H_M$ we get:

\begin{eqnarray}
\rho'_{S} &=& \PartTr{W\stateBasis{\psi}{M}\braBasis{\psi}{M}\otimes\stateBasis{\phi}{S} \braBasis{\phi}{S}}{H_M} \\
&=&  \sum_m |\braketBasis{m}{\psi}{M}|^2\, U_m\stateBasis{\phi}{S}\braBasis{\phi}{S} U_m^\dag,
\end{eqnarray}
which is identical to $\rho_{S}$ found when we performed the projective measurement followed by the conditional unitary transformation. As Alice and Bob's later measurement at $t$ only depends on the state of the subsystem $S$ (and not on the state for the subsystem $M$), we have shown that each sequence of measurement followed by conditional unitary evolution is indistinguishable from an unconditional unitary transformation, as far as Alice and Bob are concerned.

By applying this property for each sequence of measurement followed by conditional unitary transformation, we have shown the following:

\begin{myProperty}
The complete quantum mechanical description of the interactive process above can be given, for any time $t$ at which a latest measurement  can be performed by Alice and/or Bob, as a unitary transformation from the initial time $t_0$ until time $t$, followed by a POVM performed on the subsystem available to the party performing the measurement.
\end{myProperty}

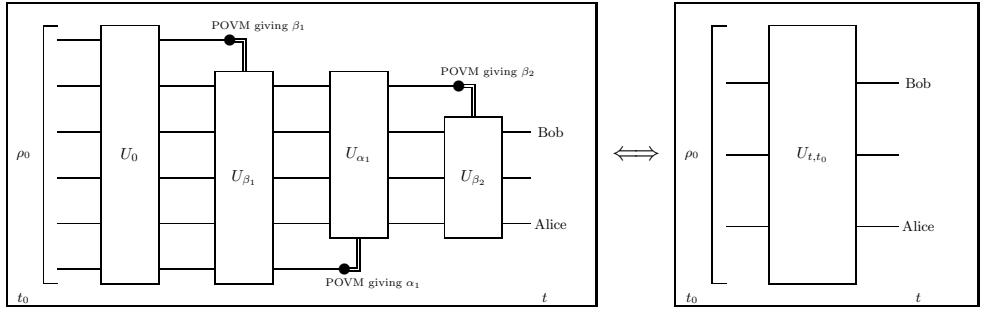
\begin{figure}[!h]
\centering
\begin{picture}(200,70)
\put(0,0){\framebox{\scalebox{0.6}{\usebox{\bigschema}}}} 
\put(132,30){\makebox(0,0){$\Longleftrightarrow$}}
\put(140,0){\framebox{\scalebox{0.6}{\usebox{\equivalent}}}} 
\end{picture}
\caption{The interactive process is not distinguishable from a single unitary transformation without any intermediary measurement }\label{FigEquivalence}
\end{figure}

This is illustrated in Figure~\ref{FigEquivalence}.

We have thus shown that, any interactive process followed by Alice and Bob --  however complex it may be -- can be described, at any point in time, as a single unitary evolution up to that time. Alice performs a measurement on the subsystem that is available to her at that time, and Bob may do the same on his subsystem. Any further transformation that is done on a subsystem that is not available to Alice cannot influence observations made by Alice.

In particular, there is no way for Alice to make sure that there was indeed any measurement done on Bob's side, despite the fact that Alice heard Bob answer to all her questions in the interactive protocol. This dilemma is similar to the well known ``philosophical zombie''~\cite{Blackmore} problem in which one wants to  determine whether a being which is behaviorally indistinguishable from a human is actually conscious or not. The quantum mechanical description of the interactive protocol (Property 1) shows that it is physically not possible for Alice to determine whether Bob is performing any measurement at all.

Interestingly, the quantum mechanical description offers an additional puzzling alternative: 
Bob may actually be experiencing something, but what he experiences may be at odds with what Alice experiences. This incoherence may go unnoticed by Alice, as what Alice hears from Bob -- which is also part of Alice's experience -- does not have to be in line with what Bob experiences on his side. 

There is no way for Alice to check in which basis Bob performs his measurement, and we have seen that Alice could experience correlation between two physical inputs while Bob sees none, or vice versa. This implies that Alice and Bob may see different causal relationships between observed events, which leads, in the end, to differences in perceived logical order of events, or more generally, to differences in perceived physical laws.

\section{Discussion}
In this note, we have proposed that two experimenters, say Alice and Bob, could have two incompatible views of the world without noticing such incompatibility. This arises because, in the view of Alice for instance, the messages from the other experimenter, Bob, are in agreement with Alice's view. Indeed, the quantum mechanical description of any interactive experiment between Alice and Bob allows the situation in which the messages received by Alice from Bob are not necessarily in line with what Bob is actually perceiving. This uncertainty arises ultimately because there is no provable relationship between the measurement bases used by Alice and the measurement bases used by Bob.

It is generally argued that the measurement basis is spontaneously determined by the unavoidable interaction of the observed quantum system with its environment~\cite{Zurek91, Zurek02}: the nature of the interaction of the quantum system with the environment would imply a ``preferred'' basis, in which the quantum state available to the observers ``decoheres'' into classical mixtures of states. 
Such theory would impose unique preferred bases for Alice and Bob, and possibly clear away the ambiguity suggested in this paper, as we could consider that the quantum information transforms into ``classical'' data long before it is communicated to Alice or observed by Bob. Unfortunately, as discussed in~\cite{Hitoshi2017}, the decoherence mechanism cannot be universal, in the sense that it cannot, in general, convert an initial state for the experimental setup into a final state that is a classical mixture of states for Alice and Bob. This is easy to be seen as we have shown that the whole experimental process can be described as a unitary evolution $U_{t_0,t}$ that maps the initial state into a final state that is (partially) observed by Alice and Bob. If you take any final state $\rho_t$ that is not a classical mixture of states for Alice and Bob, then naturally the initial state $\rho_0 = U_{t_0,t}^{-1}\rho_tU_{t_0,t}^{-1\dag}$ does not lead to a classical mixture of states.

Even if we assume an unknown additional mechanism that does select an unique basis in which the states available to Alice and Bob become classical mixture of states, there is still no theoretical reason why Alice, for instance, should perform measurement in that basis. In any case, this paper demonstrates that there is no experimental way for Alice to ascertain that Bob is performing his measurement in a given basis, as no physical protocol allows to determine what the other experimenter is actually witnessing.


\end{document}